\documentclass[lettersize,journal]{IEEEtran}
\usepackage{amsmath,amsfonts,amssymb}
\usepackage{algorithmic}
\usepackage{algorithm}
\usepackage{array}
\usepackage[caption=false,font=normalsize,labelfont=sf,textfont=sf]{subfig}
\usepackage{textcomp}
\usepackage{stfloats}
\usepackage{url}
\usepackage{eso-pic}
\AddToShipoutPictureFG*{\AtPageLowerLeft{\put(\LenToUnit{\dimexpr(\paperwidth-\textwidth)/2\relax},\LenToUnit{0.22in}){\parbox[b]{\textwidth}{\centering\scriptsize
\copyright~2026 IEEE. Personal use of this material is permitted. Permission from IEEE must be obtained for all other uses, in any current or future media, including reprinting/republishing this material for advertising or promotional purposes, creating new collective works, for resale or redistribution to servers or lists, or reuse of any copyrighted component of this work in other works.}}}}
\usepackage{verbatim}
\usepackage{graphicx}
\usepackage{cite}
\usepackage{booktabs}
\usepackage{color}
\usepackage{multirow}
\newcolumntype{C}[1]{>{\centering\arraybackslash}p{#1}}
\usepackage[colorlinks=true, linkcolor=blue, citecolor=blue, urlcolor=blue]{hyperref}
\begin{document}

\title{MetaCom: Meta Codebook Learning for Robust V2X Semantic Communications}

\author{Muge~Yu, 
        Danlan~Huang,~\IEEEmembership{Member,~IEEE,}
        Xinyi~Zhou, 
        Shuqi~Chai, 
        and~Zhiyong~Feng,~\IEEEmembership{Senior~Member,~IEEE}
        
\thanks{This work was supported by the Fundamental and Interdisciplinary Disciplines Breakthrough Plan of the Ministry of Education of China under Grant JYB2025XDXM119, the Beijing Natural Science Foundation under Grant 4262011, the National Key R\&D Program of China under Grant 2024YFC3015302, the National Natural Science Foundation of China under Grant 62321001, the Fundamental Research Funds for the Central Universities under Grant 2023RC78, and the Shenzhen Science and Technology Program under Grant RCBS20231211090816034. \textit{(Corresponding author: Danlan Huang.)}}%
\thanks{Muge Yu, Danlan Huang, Xinyi Zhou, and Zhiyong Feng are with the Key Laboratory of Universal Wireless Communications, Ministry of Education, Beijing University of Posts and Telecommunications, Beijing 100876, China (e-mail: yumuge@bupt.edu.cn; huangdl@bupt.edu.cn; zhouxinyi@bupt.edu.cn; fengzy@bupt.edu.cn).}%
\thanks{Shuqi Chai is with the Shenzhen Research Institute of Big Data, Shenzhen 518000, China (e-mail: schai@sribd.cn).}%
}

\markboth{Accepted by IEEE Communications Letters, DOI: 10.1109/LCOMM.2026.3750583}%
{Author \MakeLowercase{\textit{et al.}}: MetaCom for Robust Semantic V2X}

\maketitle

\begin{abstract}
Collaborative perception in Vehicle-to-Everything (V2X) networks faces dual challenges: strictly limited bandwidth and channel impairments. Existing static quantization schemes often suffer from feature manifold drift caused by highly non-stationary traffic environments. To address this, we propose MetaCom, a robust digital semantic communication framework featuring a Meta-Codebook Learning paradigm. First, we introduce a Differentiable Quantized Semantic Communications (DQSemCom) module trained with perturbations induced by the simulated digital modulation and channel via Gumbel-Softmax relaxation. Building on this, we design a Bi-Level Meta-Evolution Mechanism (Bi-MEM) that treats manifold drift as a sequence of rapid adaptation tasks, aligning the codebook via data-level smoothing and parameter-level meta-updates. Furthermore, we validate MetaCom under AWGN, Rayleigh, and Rician fading channels. Experimental results demonstrate that MetaCom achieves superior trade-offs between perception precision and communication overhead, outperforming state-of-the-art methods by a significant margin under low-SNR conditions.
\end{abstract}


\section{Introduction}

\IEEEPARstart{M}{ulti-agent} collaborative perception is pivotal to modern Intelligent Transportation Systems (ITS). By aggregating environmental information from adjacent vehicles and infrastructure, agents can effectively transcend the sensing limitations of occlusion and long-range detection \cite{1, 2, 3}. 

However, practical deployment faces formidable challenges. Transmitting raw, high-dimensional sensor data (e.g., LiDAR point clouds) imposes an overwhelming burden on limited Vehicle-to-Everything (V2X) bandwidth, so mainstream approaches construct sparse communication topologies and compress features. For example, Liu \textit{et al.} \cite{4} and \cite{5} introduced handshake mechanisms to achieve partial agent-level connectivity. More recently, Hu \textit{et al.} \cite{6} proposed \textit{Where2comm}, which utilizes spatial confidence maps to transmit only critical regional features. To further compress communication volume, recent works \cite{7} introduced quantization codebooks to map high-dimensional features into a compact discrete space. 

Despite these advances, two significant challenges remain unresolved. First, existing quantization frameworks predominantly rely on static codebooks trained offline. In real-world traffic scenarios, complex intersections and diverse road geometries render the distribution of semantic features highly non-stationary, leading to a critical manifold mismatch where a fixed codebook fails to cover the drifting feature space. Second, the majority of existing studies operate under idealized channel assumptions, neglecting the fading characteristic of urban V2X links. Under such constraints, channel noise distorts the transmitted semantic features, and the distortion is further amplified by downstream decoding networks \cite{8}, rendering static semantic compression systems insufficiently robust.

Recently, task-oriented Deep Joint Source-Channel Coding \cite{9} has proven effective in enhancing transmission robustness against channel variations. Meanwhile, Reptile \cite{10}, a first-order gradient-based meta-learning algorithm, enables models to rapidly adapt to new tasks by optimizing parameter initialization. In this letter, we synergize these concepts to tackle the severe feature manifold drift in dynamic V2X networks by proposing MetaCom, a robust digital semantic communication framework. The main contributions are summarized as follows:
\vspace{-0.2em}
\begin{itemize}
    \item We propose MetaCom, a digital semantic communication framework for non-stationary V2X environments, whose Differentiable Quantized Semantic Communications (DQSemCom) module optimizes discrete latent representations end-to-end for resilience against physical-layer noise.
    \item To combat feature manifold drift, we develop a Bi-Level Meta-Evolution Mechanism (Bi-MEM) that treats distribution shifts as a series of fast-adaptation tasks and keeps the codebook and encoder aligned with the drifting manifold at small overhead.
    \item Experiments show that MetaCom maintains high Average Precision (AP) across channel models at drastically reduced communication volume, outperforming existing collaborative perception methods.
\end{itemize}

\begin{figure*}[!t]
\centering
\includegraphics[width=0.8\textwidth]{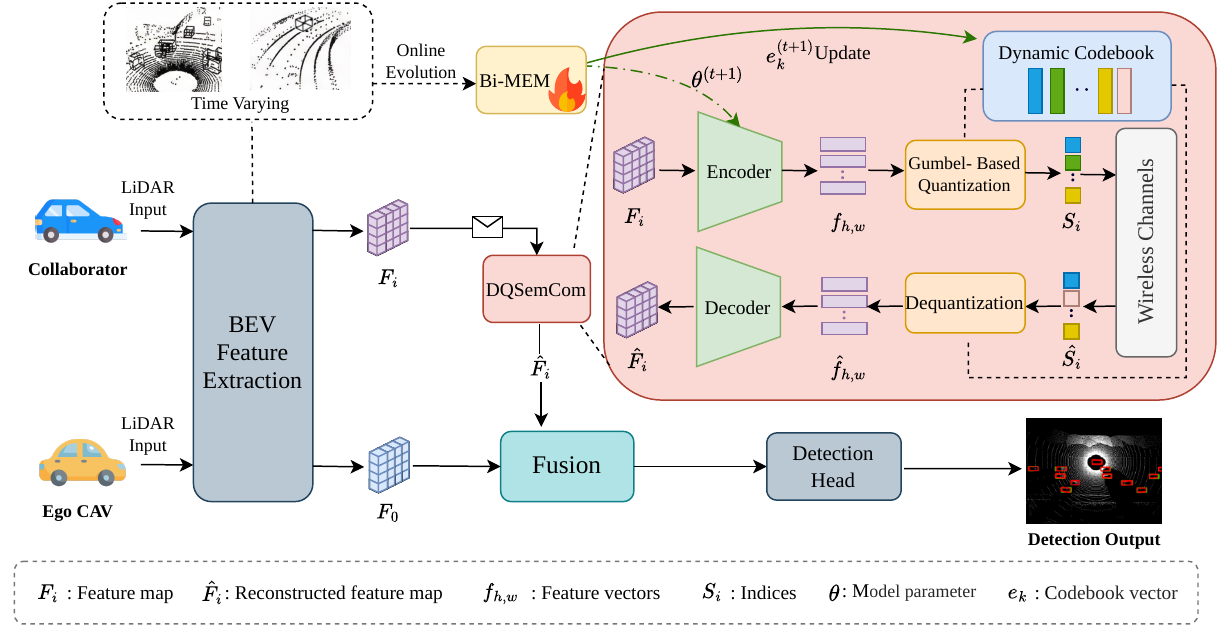} \caption{System architecture of MetaCom. CAV denotes a connected and automated vehicle. DQSemCom transmits robust discrete indices instead of high-dimensional features over noisy channels, and the codebook is updated online via Bi-MEM to follow environmental shifts.}
\label{fig:framework}
\end{figure*}

\section{System Model and Problem Formulation}

\subsection{Collaborative Perception Framework}
Consider a V2X network with an ego vehicle (index 0) and $M$ connected vehicle or infrastructure agents. At time slot $t$, agent $i \in \{0, \dots, M\}$ captures a raw LiDAR point cloud $\mathbf{P}_i^{(t)}$, and the timestamps and poses provided with each cooperative sample are used to align all features in the ego coordinate frame before fusion. The collaborative pipeline comprises three stages:
\subsubsection{Feature Extraction and Quantization} 
Each agent $i \neq 0$ extracts a Bird's-Eye-View (BEV) feature map $\mathbf{F}_i^{(t)} = E_{\theta}(\mathbf{P}_i^{(t)}) \in \mathbb{R}^{H \times W \times D}$ using a semantic encoder $E_{\theta}$. Here, $H, W$ denote spatial dimensions and $D$ is the channel size. For bandwidth efficiency, a dynamic quantizer $\mathcal{Q}$ compresses $\mathbf{F}_i^{(t)}$ into discrete symbol indices $\mathbf{S}_i^{(t)}$ via a time-varying codebook $\mathcal{C}^{(t)} = \{\mathbf{e}_k^{(t)}\}_{k=1}^K$:
\begin{equation}
    \mathbf{S}_i^{(t)}[h,w] = \mathop{\arg\max}_{k \in \{1,\dots,K\}} s\big(\mathbf{f}_{h,w}^{(t)}, \mathbf{e}_k^{(t)}\big),
\end{equation}
where $\mathbf{f}_{h,w}^{(t)}$ is the feature vector at spatial coordinates $h \in [1, H]$ and $w \in [1, W]$, $\mathbf{e}_k^{(t)}$ is the $k$-th codeword, and $s(\cdot,\cdot)$ is the similarity score defined in Sec.~III-A. This bounds the communication volume to $H \cdot W \cdot \log_2 K$ bits per frame.

\subsubsection{Wireless Transmission} 
Indices $\mathbf{S}_i^{(t)}$ are mapped to complex symbols $\mathbf{X}_i^{(t)} = \mathcal{M}(\mathbf{S}_i^{(t)})$ via digital modulation $\mathcal{M}(\cdot)$; each index is carried by one $K$-ary symbol. The received signal from agent $i$ under Additive White Gaussian Noise (AWGN) $\mathbf{N}_i^{(t)}$ is:
\begin{equation}
\label{eq:received}
    \mathbf{Y}_i^{(t)} = \mathbf{H}_{\text{eff},i}^{(t)} \odot \mathbf{X}_i^{(t)} + \mathbf{N}_i^{(t)}.
\end{equation}
where $\odot$ is the element-wise product and each entry $H_{\text{eff},i}^{(t)}$ of $\mathbf{H}_{\text{eff},i}^{(t)}$ is a flat Rician fading coefficient:
\begin{equation}
\label{eq:rician}
    H_{\text{eff},i}^{(t)} = \sqrt{\frac{\kappa}{\kappa+1}} + \sqrt{\frac{1}{\kappa+1}}\, h_i^{(t)},
\end{equation}
where $h_i^{(t)}\sim\mathcal{CN}(0,1)$ is the scattered component and $\kappa$ is the Rician factor; $\kappa\to 0$ gives Rayleigh fading and $\kappa\to\infty$ the AWGN channel. Each transmitted symbol is treated as flat-faded with an independently drawn channel realization, and the SNR denotes the average received symbol SNR.
Channel impairments cause demodulation errors, yielding a distorted index map $\hat{\mathbf{S}}_i^{(t)} = \mathcal{M}^{-1}(\mathbf{Y}_i^{(t)})$, from which the ego vehicle reconstructs $\hat{\mathbf{F}}_i^{(t)}$ by querying the synchronized codebook:
\begin{equation}
    \hat{\mathbf{F}}_i^{(t)} = \mathcal{C}^{(t)}\big(\hat{\mathbf{S}}_i^{(t)}\big).
\end{equation}

\subsubsection{Multi-Agent Fusion} 
The ego vehicle fuses its local feature $\mathbf{F}_0^{(t)}$ with the received ones via the fusion module $\Phi$ and detection head $D_{\phi}$ of \cite{6}:
\begin{equation}
    \hat{\mathbf{O}}^{(t)}(\theta, \mathcal{C}^{(t)}) = D_{\phi}\left( \Phi \left( \mathbf{F}_0^{(t)}, \hat{\mathbf{F}}_1^{(t)}, \dots, \hat{\mathbf{F}}_M^{(t)} \right) \right).
\end{equation}

\subsection{Problem Formulation}
To handle feature manifold drift under a strict bandwidth budget $B$ shared equally by the $M$ agents, our framework jointly optimizes the encoder $\theta$ and dynamic codebook $\mathcal{C}^{(t)}$. Let $\mathcal{L}_{det}$ be the detection loss against the ground-truth $\mathbf{G}^{(t)}$. The problem is to minimize the expected detection loss $\mathbb{E}[\cdot]$:
\begin{equation}
\label{eq:problem_formulation}
\begin{aligned}
& \min_{\theta, \mathcal{C}^{(t)}} \; \mathbb{E} \left[ \mathcal{L}_{det}\left(\hat{\mathbf{O}}^{(t)}(\theta, \mathcal{C}^{(t)}), \mathbf{G}^{(t)}\right) \right] \\
& \text{s.t.} \quad H \cdot W \cdot \log_2 K \le B/M \ \ \text{for every agent } i.
\end{aligned}
\end{equation}
Solving this optimization motivates the design of our scheme introduced subsequently.

\section{Proposed MetaCom Scheme}

MetaCom (Fig. \ref{fig:framework}) operates in two phases. First, an offline-trained discrete bottleneck achieves inherent robustness against channel fading via the index errors of the simulated digital link. Second, to overcome static codebook mismatches, the system updates online, treating distribution shifts as a continuous adaptation problem so that semantic consistency is kept under low bandwidth.

\subsection{Module A: Differentiable Quantized Semantic Communications (DQSemCom)}
DQSemCom projects high-dimensional BEV features into a compact discrete codebook space without breaking gradient propagation.

\subsubsection{Channel-Aware Gumbel Quantization}
 We denote $\mathbf{z}_j \in \mathbb{R}^D$ as the feature vector (i.e., flattened $\mathbf{f}_{h,w}^{(t)}$) at 1D spatial index $j \in \Omega$. We maintain a learnable codebook $\mathcal{C} = \{\mathbf{e}_k\}_{k=1}^K$. To enable gradient propagation through the discrete selection, we adopt the Gumbel-Softmax relaxation.
The selection probability $p_{j,k}$ based on the scaled dot-product similarity $s_{j,k} = \mathbf{z}_j^{\top}\mathbf{e}_k/\sqrt{D}$ between $\mathbf{z}_j$ and codeword $\mathbf{e}_k$ is computed as:
\begin{equation}
    p_{j,k} = \frac{\exp((s_{j,k} + g_k) / \tau)}{\sum_{m=1}^K \exp((s_{j,m} + g_m) / \tau)},
\end{equation}
where $g_k \sim \text{Gumbel}(0, 1)$ are i.i.d. noise samples, and $\tau$ is the temperature.
During training we use the straight-through Gumbel-Softmax sample $\bar{p}_{j,k}=\delta_{k,k_j}+p_{j,k}-sg[p_{j,k}]$ with $k_j=\arg\max_k p_{j,k}$, where $\delta_{k,k_j}$ is the one-hot indicator of $k_j$ and $sg[\cdot]$ denotes stop-gradient, so that its forward value is the one-hot vector of $k_j$ and its gradient is that of $p_{j,k}$. The index $k_j$ is passed through the same modulation, channel and demodulation chain as in \eqref{eq:received}--\eqref{eq:rician}, giving a received index $\tilde{k}_j$, and the vector delivered to the decoder is
\begin{equation}
\label{eq:index_noise}
    \hat{\mathbf{z}}_j = \sum_{k=1}^K \bar{p}_{j,k}\,\mathbf{e}_k + \big(\mathbf{e}_{\tilde{k}_j} - \mathbf{e}_{k_j}\big)=\mathbf{e}_{\tilde{k}_j},
\end{equation}
i.e., the decoder receives exactly the codeword selected by the received index, so training sees the codeword substitution that an index error causes at the operating SNR, while gradients reach $\theta$ and $\mathcal{C}$ through $\bar{p}_{j,k}$. During inference, we perform hard quantization using $k_j^* = \arg\max_k s_{j,k}$ (no Gumbel noise), transmitting only discrete indices.

\subsubsection{Sparsity-Aware Optimization}
Standard MSE loss tends to overfit the background in sparse LiDAR BEV maps. To address this, we jointly optimize the encoder parameters $\theta$ and codebook $\mathcal{C}$ via a hybrid objective prioritizing foreground features:
\begin{equation}
    \min_{\theta, \mathcal{C}} \mathcal{L}_{\text{total}} = \frac{1}{|\Omega|} \sum_{j \in \Omega} \left( w_j \| \mathbf{z}_j - \hat{\mathbf{z}}_j \|_2^2 + \lambda_{c} \| \mathbf{z}_j - sg[\mathbf{e}_{k_j^*}] \|_2^2 \right).
\end{equation}
To penalize background dominance, the weighting factor $w_j$ is set to $\alpha$ ($\alpha > 1$) for foreground regions and $1$ otherwise, where foreground denotes BEV cells containing LiDAR returns and background denotes empty cells. The second term is the commitment loss, where $\lambda_c$ is its weight; it constrains the encoder output $\mathbf{z}_j$ to align with the selected codeword $\mathbf{e}_{k_j^*}$.

\begin{figure}[t]
\centering
\includegraphics[width=0.85\columnwidth]{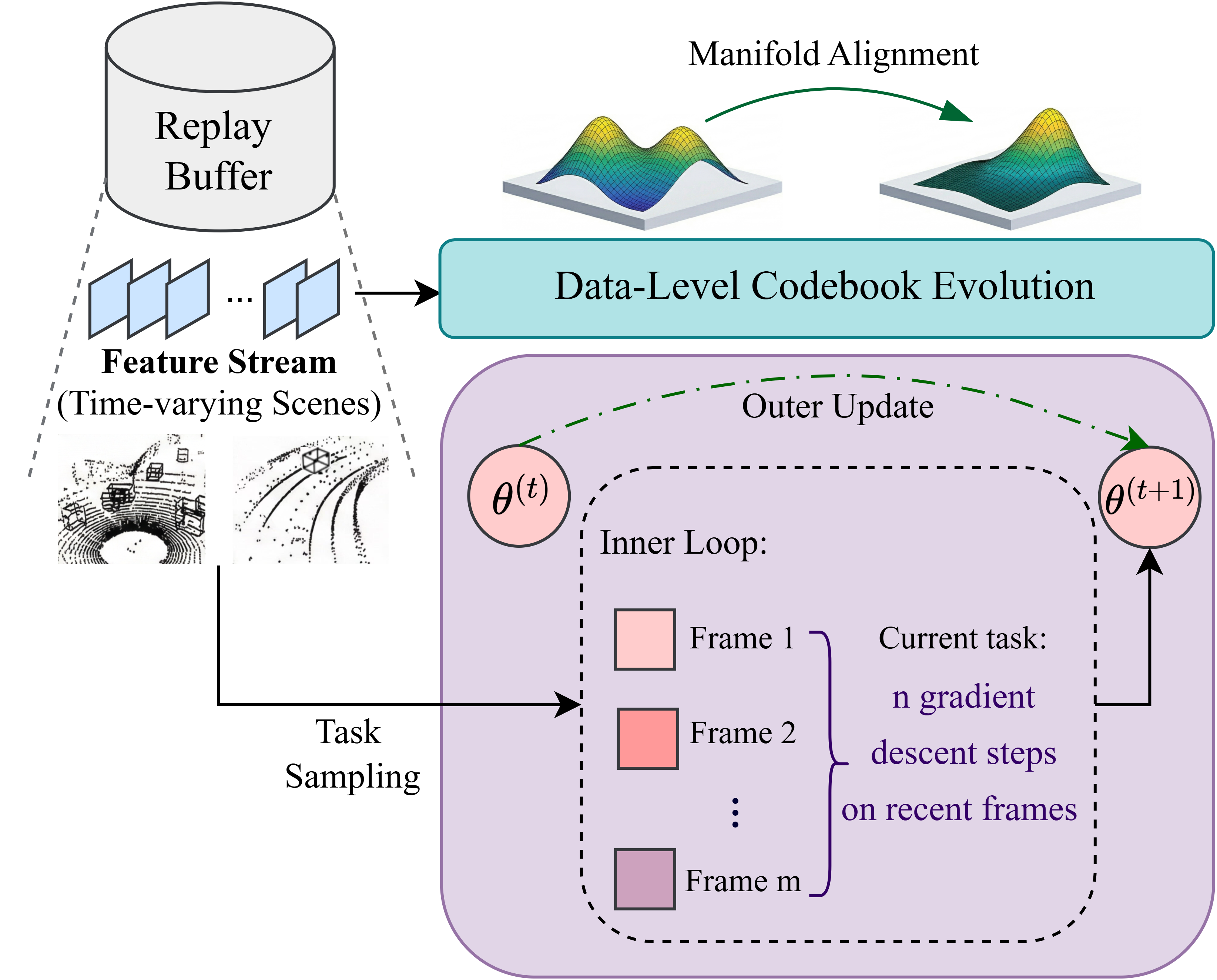}
\caption{Illustration of Bi-MEM. Level 1 updates the codebook using hard sample replay (Data Level), while Level 2 adapts the encoder via meta-learning (Parameter Level).}
\label{fig:method}
\end{figure}
\subsection{Module B: Bi-Level Meta-Evolution Mechanism (Bi-MEM)}
To effectively resolve the manifold mismatch in highly dynamic scenarios, the proposed Bi-MEM performs continuous system adaptation through two nested levels: data-level tracking and parameter-level generalization (Fig. \ref{fig:method}).

\subsubsection{Level 1: Data-Level Manifold Tracking}
As traffic scenes evolve, the optimal quantization centroids inevitably shift. We maintain a replay buffer and rank recent feature vectors by their reconstruction errors $\varepsilon_j^{(t)}=\|\mathbf{z}_j-\hat{\mathbf{z}}_j\|_2^2$. The threshold is obtained from the empirical cumulative distribution of these reconstruction errors, and hard samples are selected from those whose errors exceed the 75\% quantile. Hard samples are features that no current codeword represents well; rather than moving every codeword, Bi-MEM re-seeds one under-utilized codeword: with $u_k^{(t)}$ the EMA usage frequency of codeword $k$ and $\bar{u}^{(t)}$ its mean, an index $k^*$ is drawn from $\{k: u_k^{(t)} < 0.75\,\bar{u}^{(t)}\}$ and smoothly updated via Exponential Moving Average (EMA):
\begin{equation}
\label{eq:ema}
    \mathbf{e}_{k^*}^{(t+1)} \leftarrow \beta\mathbf{e}_{k^*}^{(t)} + (1-\beta) \bar{\mathbf{z}}_{\mathrm{hard}}^{(t)},
\end{equation}
where $\bar{\mathbf{z}}_{\mathrm{hard}}^{(t)}$ is the mean of the most recent hard samples and $\beta$ is the momentum coefficient.

\subsubsection{Level 2: Parameter-Level Meta-Adaptation}
While EMA handles geometric drift in the codebook, the encoder $\theta$ must correspondingly adapt its projection space. To achieve this, we adopt a Reptile-based meta-learning strategy. Unlike naive fine-tuning, Reptile discovers a centralized parameter initialization that is mutually close to various task-specific optima. This parameter-space proximity enables the encoder to rapidly align with new feature distributions using minimal gradient steps, while the small meta step keeps it anchored to a slowly moving initialization.
A task is the feature distribution of one scene segment (a continuous recording at one location, e.g., one intersection in DAIR-V2X); successive segments differ in traffic layout and occlusion, and the recently received features of the current segment serve as the task data. As illustrated in Fig. \ref{fig:method}, the inner loop performs $n$ gradient descent steps on the reconstruction loss of the current task, yielding a temporary parameter state $\tilde{\theta}$. Subsequently, the outer loop extracts this generalized drift direction to update the global encoder:
\begin{equation}
    \theta^{(t+1)} \leftarrow \theta^{(t)} + \gamma ( \tilde{\theta} - \theta^{(t)} ),
\end{equation}
where $\gamma$ is the meta-learning rate. The inner loop senses short-term scene variations, while the outer loop consolidates a globally generalizable initialization. This ensures the encoder adapts synchronously with the shifting codebook.

\section{Simulation Results}

\subsection{Experimental Setup}

\subsubsection{Datasets}
We evaluate MetaCom on two benchmarks: DAIR-V2X\cite{12} and OPV2V \cite{13}. DAIR-V2X is a real-world vehicle--infrastructure dataset; we use its cooperative LiDAR detection split (over 9,000 synchronized frame pairs, detection range $100m \times 40m$). OPV2V is a V2V simulation dataset with 73 scenes (11,464 frames). The datasets supply point clouds, poses and annotations; the channel impairments in \eqref{eq:received}--\eqref{eq:rician} are generated by the stochastic models at controlled SNRs. Standard splits and protocols are followed.

\subsubsection{Implementation Details}
The semantic encoder utilizes the PointPillar backbone with a voxel resolution of $(0.4m, 0.4m)$. We train the model for 30 epochs using the AdamW optimizer with an initial learning rate of $10^{-3}$ and weight decay of $10^{-2}$. All experiments are conducted on an NVIDIA GeForce RTX 3090 GPU.
For the proposed Bi-MEM, the codebook size is set to $K=64$ (6-bit indices, one 64-QAM symbol per index, $\alpha=5$), and the EMA momentum is $\beta=0.99$ to ensure stability. For the parameter-level meta-adaptation, the inner loop performs $n=5$ steps with a learning rate of $0.01$, and the outer meta-learning rate is set to $\gamma=0.001$. To simulate channel impairments, we test the system under AWGN, Rayleigh fading, and Rician fading with $\kappa=0$~dB, the mean $K$-factor of the urban NLOSv V2V link in 3GPP TR~37.885 \cite{11}; fading realizations are independent across symbols (Sec.~II-A).

\subsubsection{Evaluation Metrics}
We assess perception performance using Average Precision (AP) under typical IoU thresholds.
For communication cost, we adopt the standard Communication Volume metric defined in \cite{6}, which measures the message size in bytes on a base-2 logarithmic scale.

\subsubsection{Baselines}
Beyond classical cooperative-perception baselines, we add two communication baselines evaluated under the same channels as MetaCom. When2com, DiscoNet and Where2comm transmit their real-valued features as power-normalized analog symbols over the same channel, without quantization or coding. Trad-Com is our implemented feature-level digital pipeline (8-bit quantization, rate-1/2 LDPC modelled by its waterfall BER characteristic, and SNR-adaptive BPSK/QPSK/16QAM modulation) following the separated source--channel setting in~\cite{14}. Semantic Base is our implemented DeepJSCC-style baseline on the Where2comm pipeline, where selected intermediate features are neural-JSCC encoded, transmitted, and reconstructed as a robust neural transmission reference~\cite{15}.

\begin{figure*}[!t]
\captionsetup[subfloat]{font=footnotesize}
\centering
\subfloat[AWGN Channel]{
    \includegraphics[width=0.31\textwidth]{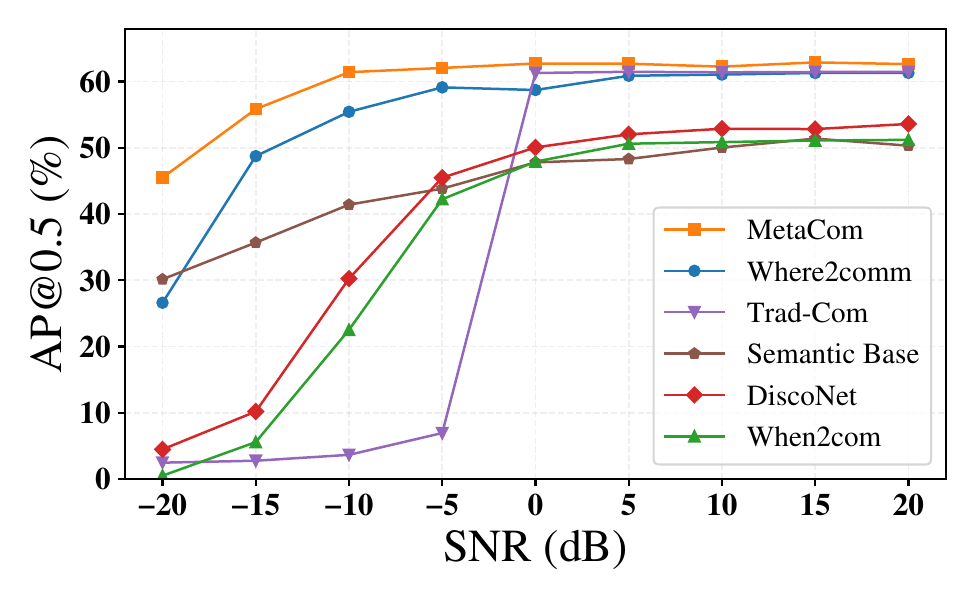}
    \label{fig:snr_awgn}
}
\hfil
\subfloat[Rayleigh Fading Channel]{
    \includegraphics[width=0.31\textwidth]{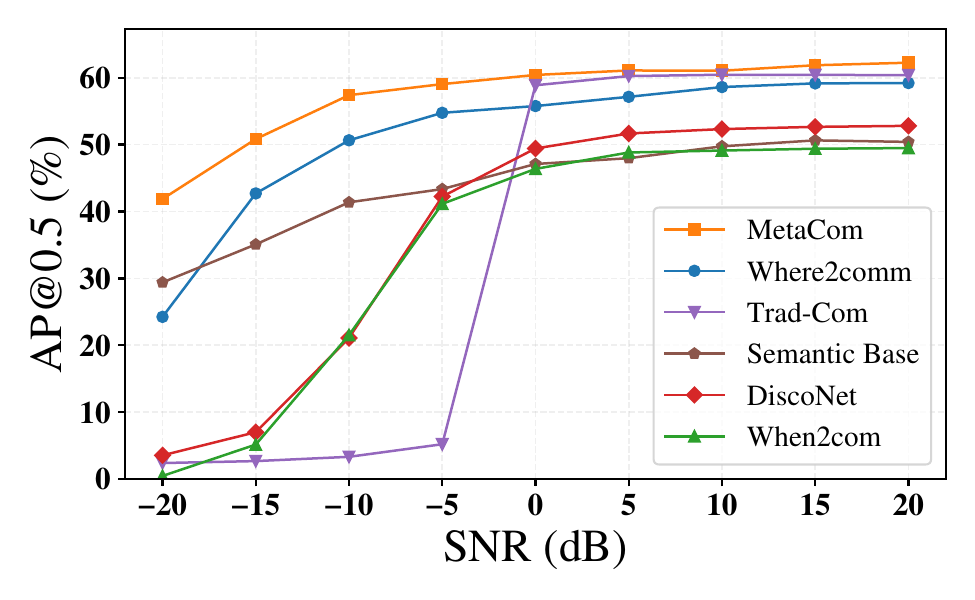}
    \label{fig:snr_rayleigh}
}
\hfil
\subfloat[Rician Fading Channel]{
    \includegraphics[width=0.31\textwidth]{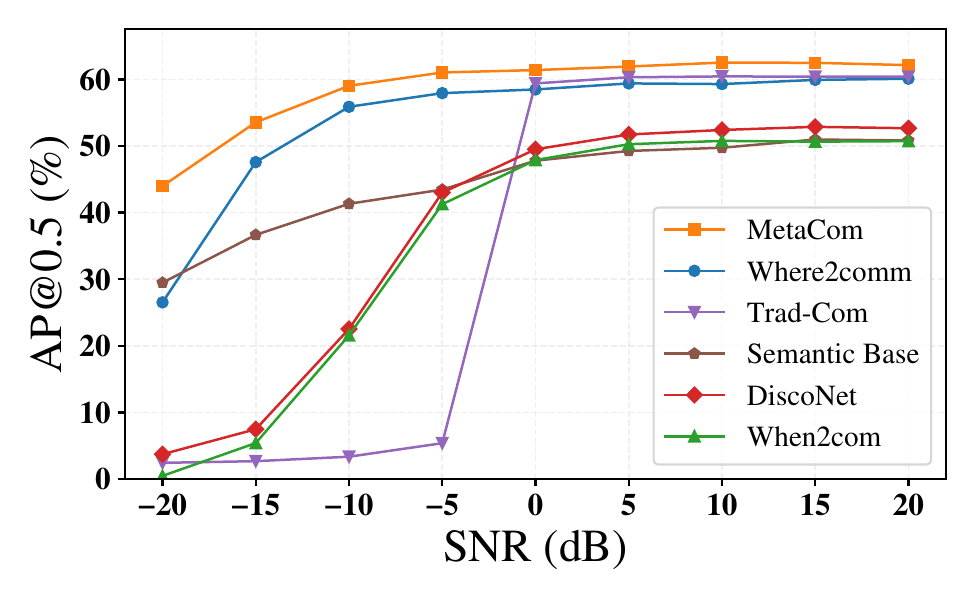}
    \label{fig:snr_rician}
}
\caption{Robustness against channel impairments (AP@0.5 vs.\ SNR) under (a) AWGN, (b) Rayleigh fading, and (c) Rician fading ($\kappa=0$~dB) channels. Trad-Com (SNR-adaptive modulation) and Semantic Base are evaluated under the same inference pipeline and channel settings as MetaCom.}
\label{fig:robustness}
\end{figure*}

\subsection{Overall Performance}
\subsubsection{Robustness over Diverse Channels}
Fig. \ref{fig:robustness} compares system resilience across three channel profiles. MetaCom consistently outperforms baselines in different channels. While analog schemes degrade severely under fading in Rayleigh (Fig. \ref{fig:robustness}b) and Rician (Fig. \ref{fig:robustness}c) channels, MetaCom maintains outstanding robustness and high AP.
\subsubsection{Qualitative Visualization}
\begin{figure*}[!t]
\centering
\includegraphics[width=0.85\textwidth]{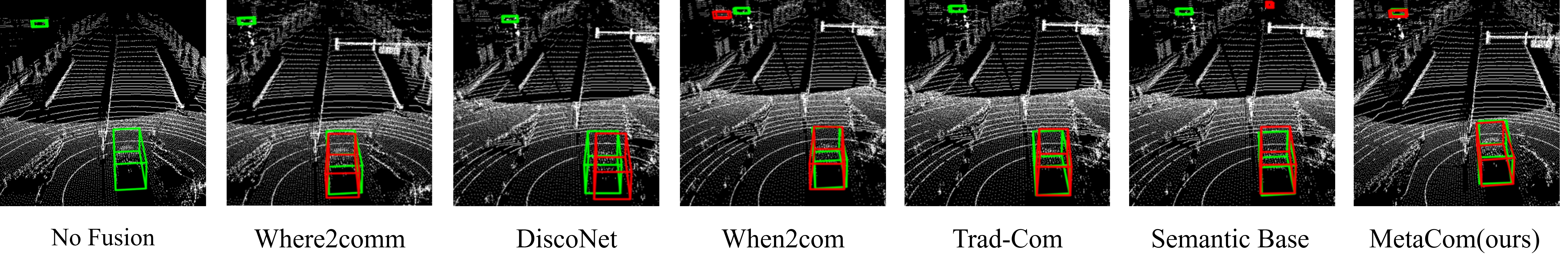}\\[-0.1em]
{\footnotesize (a) 3D Point Cloud View}\\[0.2em]
\includegraphics[width=0.85\textwidth]{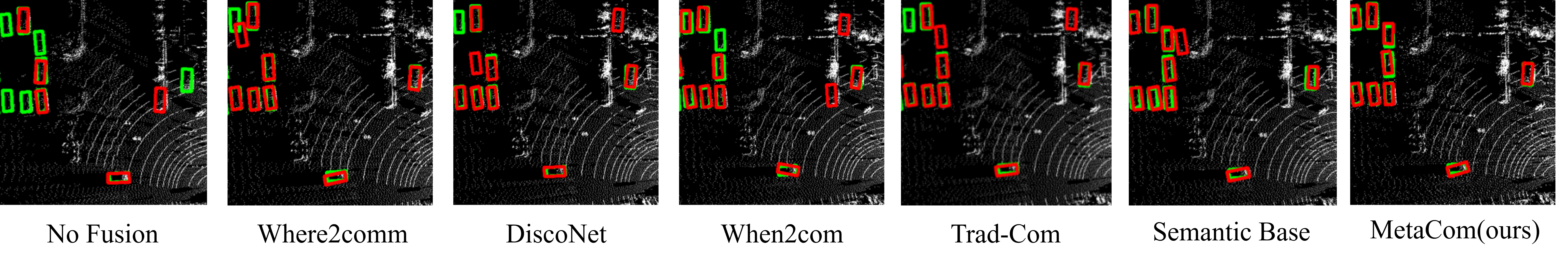}\\[-0.1em]
{\footnotesize (b) Bird's-Eye-View (BEV)}
\caption{Qualitative comparison under AWGN channel (SNR=10dB). Green boxes: Ground Truth; Red boxes: Predictions.}
\label{fig:vis}
\end{figure*}

Fig. \ref{fig:vis} visualizes detections under AWGN (SNR=10dB) in challenging scenarios with limited training samples. In both 3D (a) and BEV (b) views, baselines miss distant objects (green boxes) or produce false positives (red boxes), whereas MetaCom yields precise bounding boxes, showing its adaptability to severe channel noise and data-scarce dynamic scenes.
\subsubsection{Performance vs. Bandwidth Trade-off}
\begin{figure}[!t]
\centering
\includegraphics[width=0.9\columnwidth]{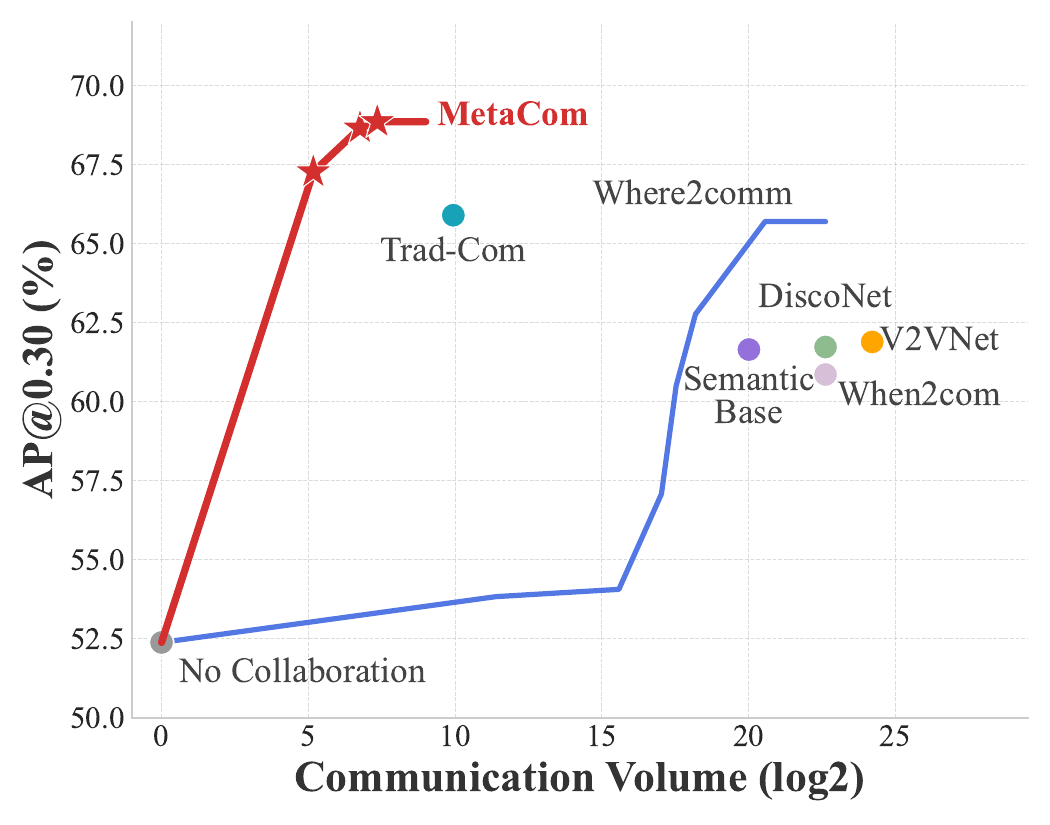}
\caption{Trade-off between detection performance (AP@0.3) and communication volume (log2 scale) on DAIR-V2X under AWGN at SNR = 20 dB.}
\label{fig:tradeoff}
\end{figure}

Fig. \ref{fig:tradeoff} illustrates MetaCom's favorable rate-distortion efficiency, where the DQSemCom module reduces data volume by orders of magnitude while keeping AP comparable to uncompressed baselines. Furthermore, Table \ref{tab:main_results} confirms consistent gains on both real and simulated data, owing to the manifold-aligned codebook and meta-initialization that adapt to new scenes without retraining from scratch.

\begin{table}[!t]
\caption{Performance comparison (AP@0.5 / 0.7, \%) on DAIR-V2X and OPV2V under AWGN at SNR = 20 dB.}
\label{tab:main_results}
\centering
\footnotesize
\setlength{\tabcolsep}{4pt}
\renewcommand{\arraystretch}{1.05}
\begin{tabular}{l|C{2.35cm}|C{2.35cm}}
\toprule
\textbf{Method} & \textbf{DAIR-V2X} & \textbf{OPV2V} \\
\midrule
No Collaboration & 50.99 / 43.38 & 68.96 / 60.89 \\
When2com~\cite{4} & 51.19 / 31.57 & 73.24 / 62.30 \\
V2VNet~\cite{1} & 59.85 / 41.01 & 80.56 / 71.75 \\
DiscoNet~\cite{2} & 53.59 / 40.59 & 79.24 / 71.67 \\
Where2comm~\cite{6} & 61.30 / 46.90 & 84.15 / 77.93 \\
Trad-Com & 61.42 / 46.67 & 85.53 / 77.91 \\
Semantic Base~\cite{15} & 51.30 / 43.66 & 83.99 / 75.65\\
IM-SemCom~\cite{14} & --- & 80.0 / 66.0 \\
\midrule
\textbf{MetaCom (Ours)} & \textbf{62.62 / 47.81} & \textbf{86.88 / 79.19} \\
\bottomrule
\end{tabular}
\end{table}

\subsection{Ablation Study}
Table \ref{tab:ablation} analyzes our modules. The first row is the shared ResNet-based BEV detection pipeline (backbone, multiscale fusion, shrink header and prediction heads), used as the complexity denominator. DQSemCom reduces the communication volume from $20.56$ to $7.35$ and raises the speed to 2.94 frame/s with an AP drop of $0.4\%$; adding Bi-MEM raises AP to 68.86\% ($+3.56\%$) at the same volume for a $13\%$ lower frame rate. FPS measures the complete prototype including data loading on one RTX 3090; the adaptation updates run between transmissions, off the critical path, and add a few milliseconds per frame amortized.
\begin{table}[!t]
\caption{Ablation and computational cost of the proposed modules.}
\label{tab:ablation}
\centering
\footnotesize
\setlength{\tabcolsep}{2.6pt}
\renewcommand{\arraystretch}{1.05}
\resizebox{\columnwidth}{!}{%
\begin{tabular}{cc|ccc|cc}
\toprule
DQ-SemCom & Bi-MEM & AP@0.30 & Comm & FPS & Trainable & FLOPs \\
 &  & (\%) &  & (frame/s) & Params (M) & (G) \\
\midrule
\multicolumn{2}{c|}{Shared BEV pipeline} & 65.70 & 20.56 & 2.83 & 9.062 & 405.34 \\
\checkmark &  & 65.30 & 7.35 & 2.94 & 9.835 & 433.14 \\
\checkmark & \checkmark & \textbf{68.86} & 7.35 & 2.56 & 9.835 & 434.74 \\
\bottomrule
\end{tabular}}
\end{table}
\section{Conclusion}
We propose MetaCom, a robust digital semantic communication framework for bandwidth-constrained collaborative perception, which combines DQSemCom for end-to-end trainable, noise-resilient transmission with Bi-MEM for online adaptation to feature drift. Experiments on DAIR-V2X and OPV2V show state-of-the-art accuracy at minimal communication overhead, outperforming existing schemes under simulated fading channels.

\end{document}